\documentclass[reprint,
 amsmath,amssymb,  aps, pre, %pra, % prb, %rmp, %prstab, %prstper, % floatfix,
]{revtex4-2}

\usepackage{graphicx}% Include figure files
\usepackage{dcolumn}% Align table columns on decimal point
\usepackage{bm}% bold math

\usepackage{graphicx}
\usepackage{float}
\usepackage[usenames,dvipsnames]{xcolor}
\newcommand{\sout}[1]{}
\newcommand{\resub}[1]{{#1}}

\newcommand{\ham}{\mathcal{H}}

\newcommand{\qq}{{\bf q }}
\newcommand{\ppp}{{\bf p }}
\newcommand{\lr}[1]{\left(#1\right)}
\newcommand*\hbz{ \hat{\bf z} }
\newcommand*\bz{ {\bf z} }
\newcommand*\Bell{\ensuremath{\boldsymbol\ell}} 
\newcommand*\Hes{\boldsymbol{\mathcal{D}}}

\definecolor{HSafron}{RGB}{232,125,30}

\begin{document}

\title{ Hamiltonian neural networks for solving equations of motion}% Force line breaks with \\

\author{Marios Mattheakis}
 \email{mariosmat@seas.harvard.edu}
%   \homepage{https://scholar.harvard.edu/marios$\_$matthaiakis}
\author{David Sondak}%
\author{Akshunna S. Dogra}
\author{Pavlos Protopapas}
 \affiliation{John A. Paulson School of Engineering and Applied Sciences, Harvard University, Cambridge, Massachusetts 02138, USA}%Lines break automatically or can be forced with \\

\date{\today}% It is always \today, today,

\begin{abstract}
There has been a wave of interest in applying {machine learning}   to study dynamical systems.  
% In particular, neural networks have been applied to solve the equations of motion, and therefore, track the evolution of a system. 
% In contrast to other applications of neural networks and machine learning, dynamical systems possess invariants such as energy, momentum, and angular momentum, depending on their underlying symmetries. 
% Traditional numerical integration methods sometimes violate these conservation laws, propagating errors in time, ultimately reducing the predictability of the method. 
We present a Hamiltonian neural network that solves  the differential equations that govern dynamical systems. 
{This is an equation-driven machine  learning method where the optimization process of the network depends solely on the predicted functions without using any ground truth data.}
The  model learns solutions that satisfy, up to an arbitrarily small error, Hamilton's equations and, therefore, conserve   the Hamiltonian invariants. 
% Once the network is optimized, the proposed architecture is considered a symplectic unit due to the introduction of an efficient parametric form of solutions. 
  The choice of an appropriate activation function drastically improves the predictability of the network. 
Moreover, an error analysis is derived and states that the numerical errors depend on the overall network performance. 
The   Hamiltonian network is then employed to solve the equations for the nonlinear oscillator and the chaotic H\'{e}non-Heiles dynamical system. In both systems, a symplectic Euler integrator requires two orders more evaluation points than the Hamiltonian network {in order} to achieve the same order of the numerical error in the predicted phase space trajectories.
\end{abstract}

\maketitle
% \mm{remove: symplectic architecture; replace unsupervised by self-supervised}

\section{Introduction}
Studying the evolution of dynamical systems has become a significant trend in scientific research. The information age has generated an exponential increase in the amount of digital data being stored, and a non-trivial fraction of these data-sets describe the evolution of dynamical systems. These include a wide range of systems, from large-scale astrophysics to nano-scale quantum physics. 
Recently, machine learning models,  and particularly  neural networks (NNs), have been used  to explore such datasets and forecast the future behavior of complex dynamical systems \cite{nanocomm2018, sapsis2018,  marios2018}, \resub{spatiotemporal chaotic behavior \cite{Ott2017, RCprl2018},}   %, ieee2019_dynSys},
classify time series \cite{rc_NN2019,lstm_NN2019},  improve turbulence models \cite{julia2016a, julia2016b, rui2019,duraisamy2019},  discover  differential equations (DEs) \cite{karniadakis2017jocp, kardiadakis2017jocp2, rudy2017scienceAdv, kutz2017data}, and find approximate solutions for those equations \cite{sinai2018, perdikaris2019}. 
In addition to the data-driven studies, equation-driven and data-free unsupervised NNs have been used to solve ordinary and partial DEs relevant to a variety of physical systems ~\cite{lagaris1998, pnas2018,spiliopoulos2018, nips2018, pra2021Quantum}.  Equation-driven networks construct analytical functions that satisfy a particular differential structure; subsequently, in the training process of such models, we do not need any ground truth data. Essentially,   the loss function solely depends on the solutions obtained by the NN  while the training process is fully data-free. This formulation results in an unsupervised learning method. We emphasize that the  proposed method  does  not  use  any  data  generated  by  traditional  numerical  solvers.
Furthermore, the universal  approximation theorem of NNs~\cite{hornik1991} states that a NN can approximate any smooth function with arbitrary accuracy. This makes NNs as a suitable approach to solving complicated problems governed by differential equations.

% Solving DEs with NNs provides some potential advantages over using traditional integrators. 
% The calculations for a NN can be efficiently implemented on parallel architectures leading to significant speed-up as pointed out by \cite{lagaris1998}. Recent hardware innovations, and in particular the wide adoption of and access to GPUs, can drastically accelerate the computation process  with minimal parallelization effort.
 %%such as GPUs naturally parallelize the calculations and thus, the NN is straight forward to get parallelized. 
 %This is a great advantage over traditional integrators where  time-parallel algorithms are challenging to develop and implement.\resub{An overview of recent advantages and challenges in parallel in time integration methods are summarized  by  \cite{parallelDE2015},  while  Ref.  \cite{parallelTraining2020}  shows that modern methods  have been invented to parallelize the time integration and can be used in deep networks for a `layer-parallel training' accelerating the network optimization.  }
%Additionally, a well known problem in numerically solving differential equations is the so-called `curse of dimensionality' that is observed as the number of the differential equations increases. It has been shown by \cite{pnas2018} that the problem of the `curse of dimensionality' can be circumvented by neural networks  and, therefore, NN differential equations solvers can  be more efficient than traditional integrators when a system is comprised of many differential equations. }

\resub{Physics-inspired and physics-informed neural networks have been widely used for solving differential equations providing some potential advantages over using traditional integrators \cite{karniadakisNatureReview2021}. 
The effectiveness of these machine learning solvers have been demonstrated by  tackling challenging problems, where traditional numerical methods become inefficient, like  solving high-dimensional PDEs \cite{pnas2018,spiliopoulos2018}, systems with moving boundary  \cite{movingBoundariesJCP2021}, and inverse problems \cite{perdikaris2019, natureSIR2021,  ANGELI2022111621}.
Solving DEs with NNs is a rapidly-growing field and new techniques are regularly proposed to advance and improve these machine learning solvers  including Monte Carlo sampling \cite{pra2021Quantum}, Fourier neural operators \cite{iclr2021fourierNet}, curriculum regularization and  sequential learning  \cite{nips2021pinns}. This work contributes to this effort by introducing a Hamiltonian structure in the NN framework that improves  the solving capability of nonlinear Hamiltonian systems.   
 The computations of a NN can be efficiently implemented on parallel architectures leading to significant speed-up  \cite{lagaris1998}. Indeed, recent hardware innovations, and in particular the wide adoption of and access to GPUs, can drastically accelerate the computation process  with minimal parallelization effort.
 %such as GPUs naturally parallelize the calculations and thus, the NN is straight forward to get parallelized. 
 This is a great advantage over traditional integrators where  time-parallel algorithms are challenging to develop and implement. An overview of  advantages and challenges in parallel in time integration methods are summarized  by  \cite{parallelDE2015},  while  Ref.  \cite{parallelTraining2020}  shows that modern methods  have been invented to parallelize the time integration and can be used in deep networks for a `layer-parallel training' accelerating the network optimization. 
  }

\resub{Data-driven Hamiltonian NNs have been proposed to impose physically informed inductive biases in the learning process. These networks  are trained faster and generalize better than regular fully connected NNs, while they  learn and respect exact conservative quantities such as the energy \cite{HNN_nips2019, chaos2019, HNN_PRE2020, vignShaan2021}.  More specifically, Greydanus et al. \cite{HNN_nips2019} introduced Hamiltonian networks with embedded the Hamiltonian formalism showing that a NN can be used to learn a Hamiltonian that describes some given temporal trajectories.  
The time derivatives and time dependence are eliminated by using Hamilton's equations and automatic differentiation resulting a time invariant energy.
Once the Hamiltonian has been learnt, predicting the motion of different initial conditions within and outside the training regime is possible by numerically solving Hamilton's equations.   Recently, this approach has been successfully applied to learn Hamiltonians,  forecast  chaotic behavior, and predict transition to chaos  \cite{HNN_PRE2020, PRR_HNN2021}.   
The  framework of the Hamiltonian NNs  is quite general and can be implemented in different  machine learning architectures like reservoir computing \cite{HamiltonianRC_PRE2021} and  graph networks   as well as, numerical  integrators can be embedded  in the network architecture  to provide   further improvement in the  long-term forecasting   \cite{Graph_hamiltonian_2019, vignShaan2021}.  Moreover, generative Hamiltonian networks have been proposed  to generate trajectories that respect certain underlying laws like energy and momentum conservation and subsequently, the generated data respect fundamental physical principles \cite{HNN_iclr2020}.
Other extensions of standard Hamiltonian networks  consider learning the dynamics of systems  in  the presence of external driven forces and dissipation. Adopting more general formulations like port-Hamiltonian, NNs   are  capable of predicting trajectories for damped  and driven time-varying dynamical systems as well as, they can efficiently uncover  underlying physical quantities hidden in data, like a stationary Hamiltonian, dissipation parameters, and external time-dependent forces \cite{pHNN2021}. 
These recent studies evidence that the learning capability of NNs  can be drastically   improved by embedding Hamiltonian formulation in the framework, nevertheless, the advantages of imposing  Hamiltonian's equations in  NNs to solve  DEs have not been studied yet. In this work we introduce and investigate a Hamiltonian NN used to solve the equations of motion of nonlinear dynamical systems.  This is an  equation-driven approach instead of data-driven model, because the form of the Hamiltonian and the initial state of a system are assumed to be known, while  ground truth trajectories (data) are not required in the training process. In other words, standard Hamiltonian networks are  learning the Hamiltonian function from  given data, whereas, our proposed model discovers trajectories that  approximately satisfy  Hamilton's equations. Subsequently, the two approaches are conceptually different, despite the fact that  Hamiltonian formulation is embedded in both networks. 
}

%{Recently, data-driven Hamiltonian NNs have been proposed to impose physically informed inductive biases in the learning process \cite{HNN_nips2019, chaos2019, HNN_PRE2020}. 
%
% \resub{\sout{In particular, Hamiltonian NNs} These networks} are able to learn and respect exact conservative quantities such as the energy. They have also been shown to train faster and generalize better than  regular fully connected NNs \cite{HNN_nips2019, chaos2019, HNN_PRE2020}.
%\resub{In Ref. \cite{HNN_nips2019} it is shown that embedding Hamilton equations in the network framework, the learning capability is significantly improved.  } Moreover, generative Hamiltonian networks are able to generate trajectories that respect certain underlying laws like energy and momentum conservation and subsequently, the generated data respect the physical principles \cite{HNN_iclr2020}.} 
%

\resub{The current work presents a {data-free} Hamiltonian neural network architecture that is used for solving DE systems. Despite the success of physics-inspired NNs in solving DEs, Hamiltonian NN solvers have not been explored yet. Subsequently, the proposed Hamiltonian NN   is an evolution of previously used data-free NNs for approximating solutions to DEs that identically satisfy boundary and initial conditions.}
We improve upon other NN DE solvers by speeding up the convergence of the network to the solution while reaping the benefits of the underlying physical properties.
We propose a NN architecture inspired by and geared towards Hamiltonian systems with  time-independent Hamiltonians. Once optimized, the NN satisfies Hamilton's equations over the entire temporal domain, directly implying the conservation of every invariant under the respective Hamiltonian flow. 
\resub{As it has been discussed in \cite{spiliopoulos2018}, calculating second derivatives using automatic differentiation is much more expensive than the calculation of first derivatives. Here, we avoid this bottleneck by solving systems of first order DEs, Hamilton's equations, instead of second order equations.}
NN solvers are conceptually different than traditional numerical solvers. Symplectic integrators are designed to conserve the energy over long time ranges. Being iterative solvers, these traditional methods accumulate errors in time and also require values  of  the  calculations  at  previous  time  points  in order to construct an approximate solution.
Traditional integrators conserve a Hamiltonian (energy) that is slightly different than the true Hamiltonian.  On the other hand, the suggested Hamiltonian NN  evaluates each time point independently and simultaneously satisfies all  the  differential  equations of a system. As a result, the Hamiltonian network  conserves the original Hamiltonian and leads to a significant reduction in any accumulated numerical errors.
\resub{Another distinct machine learning direction is the development of neural network integrators \cite{pnas2018, sinai2018}. These  hybrid models  combine traditional integrators with NNs improving the  performance in solving DEs. Our  NN solver does not belong to this class of machine learning methods since it does not require a structured mesh or embed any iteration algorithm. On the other, the proposed model suggests an alternative way to solve ODEs with neural networks without embeding conventional integrators.}
{The proposed Hamiltonian NNs consist of a more numerically precise and robust method to solve dynamical equations than standard semi-implicit schemes such as a symplectic Euler integrator. 
By sharing the network weights, choosing a trigonometric activation function, \resub{penalizing violations in energy conservation law,} and using an efficient parametric form of solutions, we show a speed-up in the convergence behavior during the optimizing process and, subsequently,  an improvement in the predictability of the network. 
Also, we show that after training the proposed  NN architecture can be considered a true and globally symplectic unit and thereby a time-invariant unit.} %and thus, time invariant unit.
%

% \resub{Neural network integrators are another direction. These networks are iterating methods outperforming other iteration methods but they still accumulate errors. These methods are more robust since the propagate the solution, building the solution on top of the previously obtained solutions. However, they accumulate errors because they satisfy a perturbed Hamiltonian.  }
%

In the rest of this study, we describe  the Hamiltonian NN architecture  that is used to approximate Hamiltonian trajectories.  An error analysis is performed and shows that the accuracy of the predicted solutions can be predefined before optimizing the network.
Then, the proposed symplectic NN is applied to solve the equations that describe the motion of a nonlinear oscillator and a two-dimensional chaotic system. \resub{We point out situations where the Hamiltonian NN solver out-performs the semi-implicit  Euler numerical method, a first order sympletic integrator. However, a comparison with higher order symplectic integrators is not presented in this work.} The network performance is demonstrated by exploring different architectures through  different parametric solutions and activation functions. \resub{Accurate long-time solutions are obtained by using a regularization term to encourage the discovery of solutions that conserve the total energy.} \resub{The experiments presented in this manuscript have been performed by using PyTorch  \cite{pytorch} and the codes are  published  on github \footnote{https://github.com/mariosmat/hamiltonianNNetODEs}.} We conclude this study with a summary of the key ideas introduced in this work, the advantages of using a Hamiltonian NN to solving DEs, and with a discussion of future plans.

\section{Hamiltonian Neural Network}
\subsection{Network architecture}

A cornerstone idea in classical mechanics is that a system's evolution can be investigated through the study of its underlying symmetries and constraints.
By the 20th century, Lagrange, Hamilton, and others had shown that the dynamics of a system is tethered to simple scalar functions, the Lagrangian and Hamiltonian functions, with multiple conservation laws and their underlying symmetries  prepackaged with these functions.  These scalar functions are then used to derive the DEs that govern the   motion of a system.  In particular, starting from the Lagrangian (the difference between kinetic and potential energy), invoking Hamilton's principle (the motion follows trajectories that minimize the  action integral), and employing techniques from the calculus of variations, the motion of a system is described by the Euler-Lagrange (E-L) equations.
In the Hamiltonian formulation, on the other hand, we start from the Hamiltonian which is a transformation of the Lagrangian and is a conservative quantity, namely it does not change in time.   This formulation  results in Hamilton's equations, which are  equivalent to the E-L equation and therefore  minimize the same action. %(in  different coordinate system). 
Hamilton's equations are a coupled set of first order DEs, whereas Lagrangian formalism provides a single set of second-order DEs. The Hamiltonian formulation possesses inherent advantages over the Lagrangian as a coupled set of first-order DEs is numerically more stable and more comfortable to solve than a single set of second-order DEs. 
Nevertheless, the resulting DEs are often analytically intractable, so engineers and scientists resort to discretization techniques to obtain solutions. 
However, the discretization procedure for solving the DEs could lead to violations of the underlying conservation laws. This issue can by cured by using   NN solvers that able to provide analytical solutions that respect the underlying principles.
Indeed, any sort of  semi-implicit method, like symplectic Euler integrator,  allows errors to %build and blow up 
{accumulate} in time. Chaotic systems in particular, are highly sensitive to such concerns and are, therefore, an ideal ground for testing the performance of the proposed  Hamiltonian NN.

We consider a physical system of many bodies that are moving in space. The  motion of those objects can be described in a $d$-dimensional configuration space which is defined by the specification of the position as a function of the time $t$ of all objects in a system. More precisely, $d$ is defined as the product of the number of bodies in a system and the number of spatial dimensions that those objects are allowed to move. In the Lagrangian formulation we are  working on the configuration space, whereas,  the Hamiltonian formalism is defined in the phase space, which consists of the position and momentum  of the objects.  Subsequently, each dimension in the configuration space associates with two degrees of freedom in the phase space. In this work, we are interested in Hamiltonian framework, therefore we consider a phase space of $D = 2d$ dimensions.
Many classical systems, from the simple pendulum to solar systems, can be described by the separable Hamiltonian form $\ham  = T  + V$, where the potential energy term $V$ depends solely on the \emph{generalized space coordinates} $\qq=(q_1, \dotsc, q_d)$, and the kinetic term $T$ depends solely on the \emph{generalized momenta} $\ppp=(p_1, \dotsc, p_d)$. Since this Hamiltonian form does not depend directly on time, systems described by it will conserve energy. Other dynamical invariants may also be inbuilt, depending upon the specific choice of the individual phase space variables and their corresponding continuous symmetries \cite{noether}. As an example, when the Hamiltonian does not directly depend on a coordinate $q_i$, the associated momentum $p_i$ is conserved  and vice versa.
For such Hamiltonian functions, the dynamics are governed by following coupled DEs, called Hamilton's  or \emph{canonical} equations:
\begin{align}
\label{eq:hamEqs}
  \dot q_i =~ \frac{\partial \ham}{\partial p_i}, \hspace{1.09cm}
  \dot p_i = -\frac{\partial \ham}{\partial q_i},
\end{align}
where dots denote time derivatives. An elegant way of expressing Hamilton's equations is the \emph{symplectic} notation. Let ${\bf z} = (q_1, \dotsc, q_d, p_1,\dotsc,p_d)^T \in {\rm I\!R}^{D}$, and  \textbf{J} be the $D\times D$ matrix 
\begin{align}
    {\bf J} =  \begin{pmatrix}
    {\bf 0} & {\bf 1} \\ {\bf -1}  & {\bf 0}     \end{pmatrix},
\end{align}
where ${\bf 0}$ and ${\bf 1}$ represent the $d \times d$ zero and unity matrix, respectively.   Then,  Hamilton's equations can be written in the  compact vector form
\begin{align}
    \label{eq:symplecticNotation}
    %\dot {\bf z}  = {\bf J} \cdot \frac{\partial \ham({\bf z})}{\partial {\bf z}}.
    \dot {\bz}  = {\bf J}\cdot \nabla_{\bz}   \ham({\bf z}),
\end{align}
where $\nabla_{\bz}   \ham({\bf z}) = {\partial \ham({\bf z})}/{\partial {\bf z}}.$
Numerical methods that evaluate Eq. (\ref{eq:symplecticNotation}) are called symplectic methods and have been widely used to calculate the long-term evolution of chaotic systems~\cite{leimkuhler1994}. 
In this work we present an alternative method based on NNs to solve Eq.  (\ref{eq:symplecticNotation}). As we will discuss below,
symplectic integrators conserve a Hamiltonian which is slightly perturbed from the original, whereas,  symplectic NNs conserve the original Hamiltonian. This is a great advantage that the proposed NN has over the symplectic integrators. 

An alternative approach to the numerically solving DEs is offered by feed-forward NNs  \cite{lagaris1998,spiliopoulos2018, karniadakisNatureReview2021}. One key advantage of such NNs over traditional numerical methods is that they seek to learn  actual functions that satisfy the DEs, rather than creating an  approximation to the real solution.
Moreover, the NN's solutions are in a closed, differentiable, and analytic form~\cite{lagaris1998}, and the calculations can  be efficiently implemented on parallel architectures leading to significant speed-ups~\cite{lagaris1998}. 
The advantage in using our proposed NN architecture is that it provides solutions that satisfies Hamilton's equations simultaneously. Thus, the dynamical invariants of a particular Hamiltonian are being  respected to the required precision, compared to the accumulation of errors that is inevitable in iterative solvers. 
To compare, we  present the  semi-implicit Euler method, which is the simplest, yet most widely used, symplectic integrator for solving Hamilton's equation. Symplectic Euler method  conserves energy up to a fluctuating error because it conserves a slightly different Hamiltonian than the original. For the separable Hamiltonian form $\ham =T(p_i) + V(q_i)$, the symplectic Euler scheme for solving the system (\ref{eq:hamEqs}) reads
\begin{align}
\label{eq:EulerScheme_q}
    q^{\lr{n+1}}_i &= q^{\lr{n}}_i +  \Delta t\frac{\partial T\left( p^{\lr{n}}_i \right) }{\partial p^{\lr{n}}_i}, \\
\label{eq:EulerScheme_p}
     p_i^{\lr{n+1}} &= p^{\lr{n}}_i -  \Delta t\frac{\partial V\left( q^{\lr{n+1}}_i \right) }{\partial q^{\lr{n+1}}_i }.
\end{align}
Here, $\Delta t$ is the time step between two sequential time points, ${\lr{n}}$ denotes the time point that is evaluated,  $q^{\lr{n}}_i = q_i(n\Delta t)$, and $p^{\lr{n}}_i = p_i(n\Delta t)$. 
Due to the iterating nature of symplectic Euler method, we read in  Eqs. (\ref{eq:EulerScheme_q}), (\ref{eq:EulerScheme_p}) that the solutions at two sequential time points are needed to evaluate Hamilton's equations at any point,  leading to numerical error in the calculation of energy, that is proportional to $\Delta t$. \resub{Similarly, higher-order iterative symplectic integrators accumulate numerical error, however, this work presents a comparison only between the solutions obtained by  the proposed NN solver and the first order  symplectic Euler integrator.}

The objective of this study is to solve Hamilton's equations (\ref{eq:symplecticNotation}) {in a certain time interval} by using NNs. Let us consider the general form of parametric solutions
\begin{align}
    \label{eq:par_sol}
     {\bf \hat z}(t) = {\bf z}(0) + f(t)  {\bf N}(t), 
\end{align}
where ${\bf \hat z}$ is the solution vector discovered by the NN, ${\bf z}(0)$ is the initial state vector, and  ${\bf N}(t) \in {\rm I\! R}^{D}$ is a vector  of $D$ outputs of a feed-forward fully connected NN. The parametric function $f(t)$ enforces the initial conditions in the parametric solutions, i.e. ${\bf \hat z}(0) = {\bf z}(0)$ when $f(0)=0$.  The network takes as a single input the time point $t_n$, where $n$ denotes the $n$-th sequential point; without  losing the generality, we consider the initial time $t_0=0$. We train the NN  by minimizing, with respect to the learning parameters of the network, 
the mean-squared error (MSE)  defined by Hamilton's equations (\ref{eq:symplecticNotation}) as:
\begin{align}
\label{eq:Loss}
	 L &= \frac{1}{K}\sum_{n=1}^K \left( {\bf \dot{\hat z}}^{\lr{n}} - {\bf J} \cdot  \nabla_{\hbz^{\lr{n}}}   \ham\left({\hbz^{\lr{n}}}\right)	 \right)^2 {+\lambda L_\text{reg}},
\end{align}
where  ${ \hbz}^{\lr{n}} = {\hbz}(t_n)$ 
and $K$ is  the  total number  of the input time points used for the network optimization. {The term $L_\text{reg}$ can be any regularization term where $\lambda$ is the regularization parameter. 
We have found that for long time predictions it is efficient to use  a regularization  term that penalizes violations of the energy conservation law. Given the initial state and corresponding energy, $E_{0}$, of a system, a convenient regularization term is}
%Specifically, since the initial state of a system and thus, the initial energy $E_0$, are known an efficient  regularization term is}
\begin{align}
\label{eq:regLoss}
{
 L_\text{reg} =  \frac{1}{K}\sum_{n=1}^K \left[ \left(\ham\left({\bf \hat z}^{\lr{n}}  \right) - E_0 \right)^2 \right].
}
 \end{align}
 %{where the first term in the summation denotes the network prediction for the energy at $t_n$.
{For long-time prediction, the use of  the regularization loss~\eqref{eq:regLoss}  stabilizes the predicted trajectory at the correct energy level  and can result in faster network convergence. In the present work, results are presented for $\lambda=0$ unless otherwise specified.}
%Nevertheless, in the work we want to emphasize in the Hamiltonian part of the loss function of Eq. (\ref{eq:Loss}), hence in the most part of the following study we consider $\lambda=0$. We consider non-zero regularization term only in long-term prediction.}

\begin{figure}
    \centering
    \includegraphics[scale=0.3]{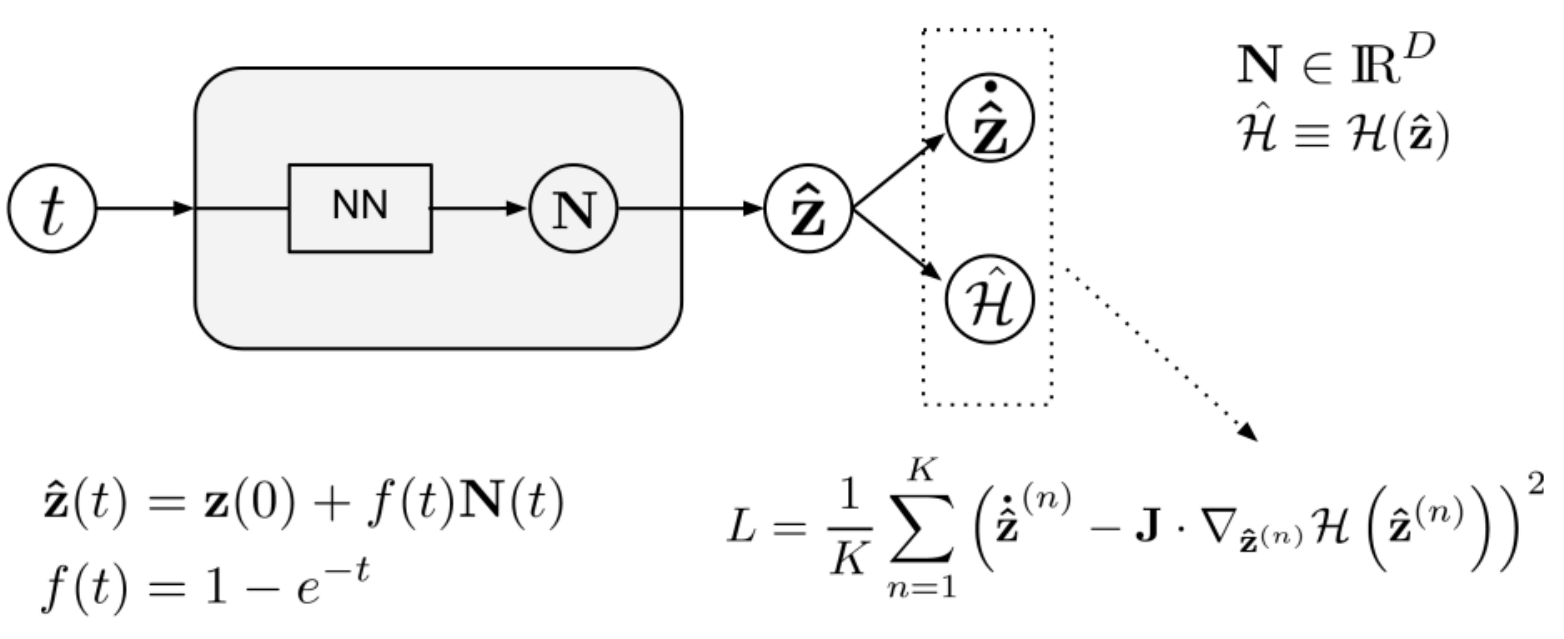}
    \caption{Hamiltonian architecture with parametrization  ${\bf \hat z}(t)$ used in the loss function $L$; $\ham$ is the Hamiltonian and $f(t)$ imposes the initial conditions to ${\bf \hat z}(t)$; $K$ is the number of the training points and $(n)$ indicates each time point. } 
    \label{fig:symNet}
\end{figure}

The time derivatives are obtained by using automatic differentiation  {that computationally costs one  back-propagation through the entire network} \cite{pytorch}. We first generate equally-spaced time points $t_n$ in the {training} time interval $[0, T]$. Then, we randomly perturb  these points in each epoch as: $t_n \rightarrow t_n + \epsilon$ where $\epsilon$ is a random  term obtained by a normal distribution \cite{spiliopoulos2018}.  This trick improves the  network predictability as  it is effectively trained over a continuous time interval. In addition, perturbing the training points in every epoch employs the stochastic gradient descent (SGD) method, and thus it  assists the optimizer to escape from local minima in the loss function. Perturbing the points in every epoch means that we perturb the loss function and, subsequently, the local minima are dynamically moving. In  the context of  SGD,  each epoch is considered as a mini-batch while   all the epochs consist the whole batch for the training set. Minimizing the loss function   in Eq. (\ref{eq:Loss})  yields solutions that identically respect the symplectic structure of Eq. (\ref{eq:symplecticNotation}) and accordingly, every dynamical invariant of the Hamiltonian flow is respected too. \resub{We point out that the NN solutions are of high accuracy when the NN is evaluated in the training interval $[0,T]$  but the error rapidly increases outside of the training interval.} \resub{\sout{As shown by  \cite{lagaris1998}, for $t<0$ or $t>T$ the extrapolation error remains low for time points close the range $[0,T]$ but it rapidly increases outside of the training interval. That happens because we employ a feed-forward NN which is not able to learn any temporal correlation since it does not have any internal memory. The use of recurrent  NNs potentially can improve  the Hamiltonian NNs in terms of extrapolation far away of the training range. However, such considerations are outside the scope of the current study.}} The proposed Hamiltonian NN architecture is graphically demonstrated in Fig. \ref{fig:symNet}. \resub{It is worth noting that the proposed network of Fig.  \ref{fig:symNet} has a different architecture than one used  in standard Hamiltonian NNs \cite{HNN_nips2019}. Our network takes  $t_n$ as input and returns $\hbz^{\lr{n}}$, whereas, the input in the standard Hamiltonian networks is $\bz^{\lr{n}}$ and the output is $\hbz^{\lr{n+1}}$.}
%coordinate {\bf \hat z}^{\lr{n}}   as input our network takes time as input whereas standard }

{}

\resub{A crucial role in the performance of the NN is played by $f(t)$. 
A standard choice to enforce initial conditions is $f(t)=t$, which satisfies $f(0)=0$  \cite{lagaris1998}.}  However, this is an unbounded  function that adds further difficulty when $t$ becomes large. Specifically, {for the NN outputs after enough epochs, Eq. (\ref{eq:par_sol}) states that  ${ \bf N} =  ({\bf \hat z-z}(0))/t$. As   $t$ increases the ${ \bf N}$ tends to zero}, which  affects negatively  the network predictability in large time scales. To remedy this inefficiency  we propose the parametric function 
\begin{align}
    f(t) = 1-e^{-t} \label{eq:f_param},
\end{align}
which is a smooth, bounded function, with $f(0)=0$. Later, we show that the specific choice of parametric function drastically improves the predictability of the NN solver.
Interestingly enough, the fact that   $f(t)$ rapidly tends to $1$  implies that the proposed  architecture consists a symplectic  NN. In particular, {for $\lambda=0$ and} at the limit $L \rightarrow 0$ Eq. (\ref{eq:Loss}) yields ${\bf \dot{\hat z}} = {\bf J} \cdot \nabla_{\hbz} \ham(\hbz)$, and as $t\to\infty$, we have ${\mathbf{\hat z}} ={\bf z}(0)+ \mathbf{N}$. Considering the aforementioned two limits and performing the linear transformation ${\bf N} \to {\bf N -z}(0)$ we obtain:
\begin{align}
    \label{eq:symplecticNN}
     \dot {\bf N}  = {\bf J} \cdot \nabla_{\bf N}  \ham({\bf N}),
\end{align}
which  indicates that the proposed architecture comprises a symplectic NN that states that the function $\ham({\bf N})$  is  time invariant.

\resub{A substantial advance that our method suggests is the energy regularization of Eq. (\ref{eq:regLoss}). Because the absence of a iteration learning, the NN solver does not build the solutions using predictions from previous steps. 
%Since the proposed NN solver does not embed an iteration method, it does not build the solution using solutions from previous time steps. 
As a result, it tends to forget the initial state of the system and thus, in long time solutions,   energy leaking might be  observed resulting to error accumulation. This issue becomes crucial in long-time solutions reducing the ability of solving nonlinear and especially chaotic systems of ODEs. The regularization loss of Eq. (\ref{eq:regLoss}) stabilizes the predicted trajectories at the correct energy yielding a robust solver.}
\resub{\sout{An important distinction between the present and previous method presented in ~\cite{lagaris1998} is that the latter proposes one NN per DE, and therefore, $D$ single-output networks are required. This is conceptually different from our approach, where we suggest one NN with $D$ outputs. By using a single network, the individual outputs share all the weights except those in the output layer, allowing correlations between the outputs. Hamilton's equations are indeed correlated and thus, by sharing the weights assists the network to discover these co-dependencies. Consequently, we obtain the same complexity with fewer learning parameters, yielding a more robust and efficient network.
We point out a set of substantial differences between our study and Lagaris' et al. work ~\cite{lagaris1998}: the depth of the NN and the choice of the activation function. In ~\cite{lagaris1998}, the network architecture is restricted to one hidden layer, uses sigmoid activation functions, and the derivatives used for the back-propagation are calculated analytically. In the current work, we use  automatic differentiation to calculate the derivatives  \cite{pytorch}. Subsequently, we were able to use any activation function and arbitrarily many hidden layers, which leads to greater control over the complexity of the network. }}
\resub{Another important innovation of this work is the choice of the activation function. It has been shown that NN with  trigonometric activation functions can  learn  periodic behavior from data outperforming networks that use common activations like Relu and Sigmoid \cite{sinArxiv}. We adopt this approach and choose the trigonometric $\sin(\cdot)$ as the activation function.  Empirical results presented later through numerical experiments indicate that $\sin(\cdot)$ activation outperforms sigmoid in solving ODEs for Hamiltonian systems.} 
%and show that the NN converges to the solutions with less training iterations than using the sigmoid activation.  
%
\resub{\sout{Using $\sin(\cdot)$ as the activation function permits the solution to be expressed on a basis that has global support similar to the Fourier series. The derivative of  $\sin(\cdot)$ has more global support than the derivatives of the traditional activation functions, such as those from the sigmoid family, hence  $\sin(\cdot)$ are a more  expressive activation function.}}

\subsection{Error Analysis}
We seek to provide a rough bound on the error in the solution based on the maximum value of the loss function.  To begin, 
%{we consider $\lambda=0$  and} 
note that Eq.~\eqref{eq:Loss} can be written as
%$L= \sum_{n}{\Bell_{n}^2}/K$   where 
{
\begin{align}
  L = \frac{1}{K}\sum_{n=1}^{K}{\lr{\Bell_{n}^{2} + \lambda\Bell_{\text{reg},n}^{2}}}
\end{align}
where
\begin{align}
    \label{eq:partial_loss}
    \Bell_{n} &=  {\bf \dot{\hbz}}^{\lr{n}} - {\bf J} \cdot \nabla_{\hbz^{\lr{n}}}\ham\left(\hbz^{\lr{n}} \right), \\
    \Bell_{\text{reg},n} &= \ham\lr{\hbz^{\lr{n}}} - E_{0},
\end{align}
}
{
and $\Bell_{n}, \Bell_{\text{reg},n} \in \mathbb{R}^{K}$ are vectors containing the respective loss components 
}
%and $\Bell_{n}, \Bell_{\text{reg},n} \in \mathbb{R}^{K}$ are vectors containing the respective loss components 
%%that contains 
%%is a vector $\Bell_n=\left(\ell_{n,1}, \dotsc, \ell_{n,D} \right)$ 
%%all the loss components 
%}
at some arbitrary time point $t_n$. Since $L$ is the loss function for the NN, averaged over time, $\Bell^2_n$ can be considered the {instantaneous} loss at the $n^{th}$ time point {when $\lambda=0$}.
Let $\delta\bz= \bz - \hbz $ be the error between   the true solution and the NN solution.  Expanding the Hamiltonian $\ham\left(\bz\right)=\ham\left(\hbz + \delta \bz \right)$ in a Taylor series about $\hbz$ and keeping up to quadratic  terms yields:
\begin{align}
 \label{eq:expH}
     \ham\left(\bz  \right) \approx \ham\left(\hbz\right) + \left(\nabla_{\bz} \ham\left(\bz \right)\right)_{\hbz}\delta \bz + \frac{1}{2} \left(\Hes_{\bz} \ham\left( \bz \right)\right)_{\hbz}\delta \bz^2,
\end{align}
where $\Hes_{{\bf{z}}}$ is the Hessian matrix.  Taking the gradient of Eq.~\eqref{eq:expH} with respect to $\bz$ and rearranging terms gives,
\begin{align}
 \label{eq:expHz}
     \left(\nabla_\bz \ham\left(\bz  \right)\right)_{\hbz} \approx  \nabla_{\bz}\ham\left(\bz \right)  - \left(\Hes_{\bz} \ham\left(\bz \right)\right)_{\hbz} \delta \bz.
\end{align}
%
% %
We note that for  Hamiltonians with quadratic dependence on $\bz$, the quadratic expansion (\ref{eq:expH}) is exact because higher order terms  vanish. In addition, the second order in $\delta \bz$ is the smallest order still large enough to  not be canceled when we move to substitute Eq.  (\ref{eq:expHz}) into Eq. (\ref{eq:partial_loss}).
Nevertheless, the derivation can be extended to include higher order terms in a straightforward manner.  In what follows, we drop the superscript $\left(n\right)$ for clarity of presentation.  Substituting the  Taylor series expansion (\ref{eq:expHz}) into~\eqref{eq:partial_loss} and invoking~\eqref{eq:symplecticNotation} results in,
\begin{align}
    \label{eq:ell}
    \Bell \approx   {\bf J} \cdot [\left(\Hes_{\bz} \ham\left(\bz\right)\right)_{\hbz}\delta{\bz}]   -  \dot{\delta \bz}.
\end{align}
Inspecting the vector DE (\ref{eq:ell}) we observe that its components comprise a closed differential system for the error $\delta z_i$ in each predicted trajectory $\hat z_i$. Solving this differential system with initial condition $\delta \bz(0) = 0$, as it is dictated by the parameterization (\ref{eq:par_sol}), we can compute how the errors propagate in time.  However, this requires knowledge of the loss components of $\Bell(t)$  and thus, such an analysis can be performed only  after we have trained the network.

On the other hand, we can derive a bound on the size of $\delta \bz$ without having exact knowledge of $\Bell(t)$ by constructing a worst case scenario.  We want to establish a relationship between $\Bell$ and $\delta \bz$, such that it determines when to stop the network training in order to get solutions with better than a certain accuracy. 
Let $\displaystyle \ell_\text{max}^2 = \max_{t}(\Bell^2 { + \lambda\Bell_{\text{reg}}^{2}})$ represent the largest instantaneous loss that the neural network will have after being trained.  In the following analysis, we denote the $2-$ norm by $\|\cdot\|$.  We have,
\begin{align}
    \ell_\text{max}^2 &\geq {\|\Bell\|^{2} + \lambda\|\Bell_{\text{reg}}\|^{2}} \nonumber \\
                      &{\geq \|\Bell\|^{2}} \nonumber \\
                      &{=} \|\dot{\delta \bz} - {\bf J} \cdot \left(\Hes_{\bz} \ham\left(\bz\right)\right)_{\hbz}\delta{\bz}   \|^{2} \nonumber \\
                      &\geq \left|\|\dot{\delta \bz}\| -  \|\left({\bf J}\cdot\Hes_{\bz} \ham\left(\bz\right)\right)_{\hbz}\delta{\bz}\| \right|^{2} \nonumber \\
                    %   &= \|\dot{\delta \bz}\|^{2} - 2\|\dot{\delta \bz}\|\|\left({\bf J}\cdot\Hes_{\bz} \ham\left(\bz\right)\right)_{\hbz}\delta{\bz}\| + \|\left({\bf J}\cdot\Hes_{\bz} \ham\left(\bz\right)\right)_{\hbz}\delta{\bz}\|^{2} \nonumber \\
                      &= \|\dot{\delta \bz}\|^{2} - 2\|\dot{\delta \bz}\|\|\left({\bf J}\cdot\Hes_{\bz} \ham\left(\bz\right)\right)_{\hbz}\delta{\bz}\| \nonumber 
                      \\ & \quad \hspace{2cm} + \|\left({\bf J}\cdot\Hes_{\bz} \ham\left(\bz\right)\right)_{\hbz}\delta{\bz}\|^{2} \nonumber \\
                      &\geq \|\dot{\delta \bz}\|^{2} - 2\|\dot{\delta \bz}\|\|\left({\bf J}\cdot\Hes_{\bz} \ham\left(\bz\right)\right)_{\hbz}\delta{\bz}\| + \left(\sigma_{\text{min}}\|\delta{\bz}\|\right)^{2} \label{eq:bounds1},
\end{align}
where $\sigma_{\text{min}}$ is the minimum singular value of $\left(\Hes_{\bz} \ham\left(\bz\right)\right)_{\hbz}$.  The last line in the above expression~\eqref{eq:bounds1} can be obtained by considering the quantity $\|Ax\|$ and using the singular value decomposition on $A$ to show that $\|Ax\| \geq \sigma_{\text{min}}\|x\|$.  Rearranging terms leads to,
\begin{align}
\sigma_{\text{min}}^{2}\|\delta{\bz}\|^{2} &\leq \ell_\text{max}^2 - \|\dot{\delta \bz}\|^{2} + 2\|\dot{\delta \bz}\|\|\left({\bf J}\cdot\Hes_{\bz} \ham\left(\bz\right)\right)_{\hbz}\delta{\bz}\| \nonumber  \\ 
    &\leq \ell_\text{max}^2 - \|\dot{\delta \bz}\|^{2} + 2\|\dot{\delta \bz}\|\|\left({\bf J}\cdot\Hes_{\bz} \ham\left(\bz\right)\right)_{\hbz}\|\|\delta{\bz}\| \nonumber \\
    \Rightarrow \sigma_{\text{min}}^{2}\|\delta{\bz}\|^{2} &- 2\|\dot{\delta \bz}\|\|\left({\bf J}\cdot\Hes_{\bz} \ham\left(\bz\right)\right)_{\hbz}\|\|\delta{\bz}\| \leq \ell_\text{max}^2 - \|\dot{\delta \bz}\|^{2}. \label{eq:dzIneq}
\end{align}
Solving the  quadratic inequality (\ref{eq:dzIneq}) for $\|\delta{\bz}\|$ yields,
\begin{align}
    \|\delta{\bz}\| &\leq \dfrac{\|\dot{\delta \bz}\|\|\left({\bf J}\cdot\Hes_{\bz} \ham\left(\bz\right)\right)_{\hbz}\|}{\sigma_{\text{min}}^{2}}  
\nonumber \\    
      +\dfrac{1}{\sigma_{\text{min}}^{2}} & \left[\sigma_{\text{min}}^{2}\ell_\text{max}^2 - \|\dot{\delta \bz}\|^{2}\left(\sigma_{\text{min}}^{2} - \|\left({\bf J}\cdot\Hes_{\bz} \ham\left(\bz\right)\right)_{\hbz}\|^{2}\right)\right]^{1/2}. \label{eq:dz_bound}
\end{align}
 Now consider a single component of the error, $\delta z_{i}$.  The largest value $\delta z_{i}$ can take occurs when $\delta z_{i}\neq 0$ and $\delta z_{j} = 0$ for $j\neq i$.  That is, for a fixed error, all of the error is concentrated in a single component.  In this case, $\|\delta \bz\|^{2} = \delta z_{i}^{2}$.  If $\delta z_{i}^{2}$ is maximized at a value $t_{\text{max}}$, then $\dot{\left(\delta z_{i}^{2}\right)} = 0$ at $t_{\text{max}}$.  Therefore, $\delta z_{i}\dot{\delta z_{i}} = 0 \Rightarrow \dot{\delta z_{i}} = 0$.  Using this in~\eqref{eq:dz_bound} provides,
 \begin{align}
     \label{eq:dz}
     \|\delta z_{i}\| \leq \frac{\ell_{\text{max}}}{\sigma_\text{min}}.
 \end{align}
\resub{We point out that at the boundary  $t=0$, the error and its derivatives are exactly zero since the initial conditions are identically satisfied through the parametrization of Eq. (\ref{eq:par_sol}). Furthermore, the  assumption that all the error is  concentrated  at a single component $z_i$ implies that $\delta z_j$ and $\dot{\delta z_j}$  are zero functions for $j\neq i$. This  strong assumption simplifies Eq. (\ref{eq:dz_bound}) yielding   the upper error bound of Eq. (\ref{eq:dz}).}

If a NN is trained such that the loss function has a maximum value of $\ell_\text{max}$, then the maximum error that any component of the solution can take is bounded by~\eqref{eq:dz}.
In other words, we can choose in advance an accuracy for the solutions and use the relationship~\eqref{eq:dz} to calculate the  $\ell_\text{max}$, which, therefore, will determine when we have to stop training the network  ensuring the desirable accuracy. The $\sigma_\text{min}$ can be calculated due to the training process since, in the most general case, it is a function of the solutions.
Moreover, the expressions (\ref{eq:ell}) and (\ref{eq:dz}) state that $|{\bf \delta z}|$ depends on the general network performance  and not only on the number of the time points used in the training process, which is the case of numerical integrators. That happens because  the number of training points is not the only parameter that determines the value of the loss function. For example, fixing the number of the training points while increasing the number of hidden layers or neurons yields better performance that corresponds to a smaller $\ell_\text{max}$.
In summary, once the Hamiltonian NN is optimized,  Eq. (\ref{eq:ell}) can be used to calculate the error propagation. On the other hand, we can decide the accuracy of the solutions before the optimization by using  Eq. (\ref{eq:dz})  to define the $\ell_\text{max}$ that determines when to stop  training the network.

\section{Experiments}
\subsection{Nonlinear Oscillator} 

As a concrete example, we consider the  one dimensional  nonlinear (an-harmonic) oscillator  with  Hamiltonian 
\begin{align}
    \label{eq:Hosc}
    \ham= \frac{p^2}{2} + \frac{x^2}{2}  + \frac{x^4}{4},
\end{align}
where the natural frequency and the mass of the oscillator are considered to be unity. The Hamiltonian (\ref{eq:Hosc}) corresponds to the total energy $E$ of the system, and the associated equations  of motion  read (Eq. \ref{eq:hamEqs}):
\begin{align}
    \label{eq:eqMotOsc}
    \dot x = p, \quad \quad \dot p = -(x+x^3).
\end{align}
In what follows, we use the symplectic NN architecture to solve the above nonlinear Hamiltonian system and compare the NN solutions with those obtained by symplectic Euler integrator. It results that the  symplectic Euler method requires two orders more evaluation time points than the NN to reach the same numerical error.
   We also explore the efficiency of the network for different activation  and  parametric functions. 

\begin{figure}
    \centering
    \includegraphics[scale=0.25]{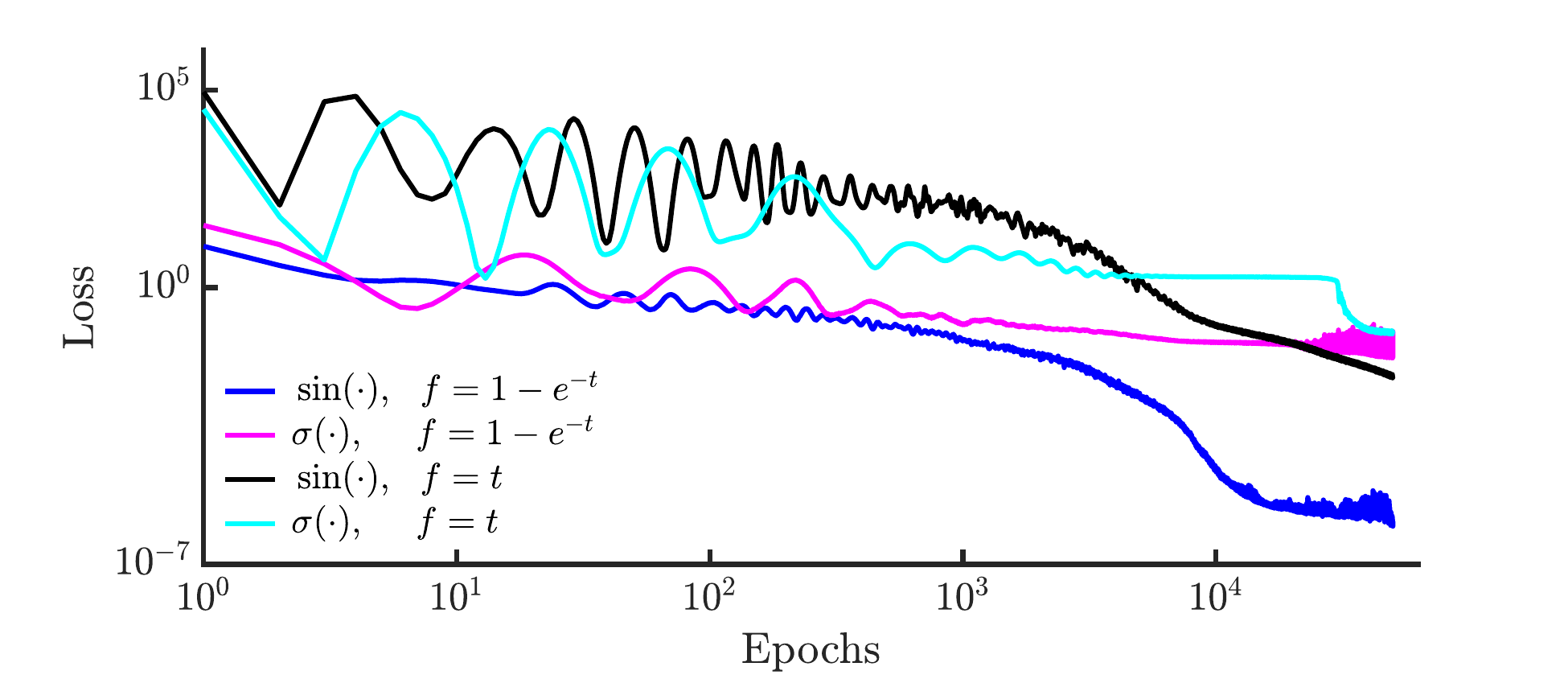}
    \caption{Hamiltonian NN solves the equations of the nonlinear oscillator system. Color lines represent the loss function in log-scale during the training  for different combinations of activation and parametric functions $f$ shown in legend.}
    \label{fig:NLosc_loss}
\end{figure}

The phase space of the oscillator consists of two degrees of freedom with ${\bf z}=(x,p)^T$. Accordingly, we utilize a feed-forward NN with two outputs ${\bf N} = (N_1, N_2)^T$  used to parametrize the approximate solutions ${\bf \hat z}=(\hat x, \hat p)^T$ according to Eq. (\ref{eq:par_sol}). The loss function is defined by  Eqs. (\ref{eq:eqMotOsc}) and according to Eq. (\ref{eq:Loss}) as: 
\begin{align}
    \label{eq:Loss_osc}
    L =  \frac{1}{K}\sum_{n=1}^K \left[ \left( \dot{\hat x}^{(n)} - \hat p^{(n)}  \right)^2+\left(\dot{\hat p}^{(n)} + \hat x^{(n)} + \left(\hat{x}^{(n)}\right)^3\right)^2\right].
\end{align}
We initialize a grid with $K=200$ time points equally spaced in the time interval $t=[0, 4\pi]$. At the beginning of each epoch, we perturb all the time points by using a random term obtained by a normal distribution with zero mean and a standard deviation of $0.06\pi$. The initial state is chosen to be $(x_0,p_0)=(1.3,1.0)$, corresponding to the total initial energy $E_0=2.06$; in this energy, the motion deviates from the  behavior of the simple harmonic oscillator. The NN consists of two hidden layers with $50$ neurons per hidden layer, and is being trained for $5 \cdot 10^4$ epochs by using Adam optimizer \cite{adam} with a learning rate of $8 \cdot 10^{-3}$. We perform four independent numerical experiments that correspond to different NN designs,  namely   for the combinations of sigmoid $\sigma(\cdot)$ and trigonometric $\sin(\cdot)$ activation functions,  and for the parametric functions  $f(t)=t$ and $f(t)=1-e^{-t}$.
Figure \ref{fig:NLosc_loss}  demonstrates in logarithmic scale the loss function  (\ref{eq:Loss_osc})  during the training; each color represents one of the  the four distinguished  cases of architectures according to the legend. We highlight that the loss function of  our proposed design (blue line) converges faster than the other models.

\begin{figure} 
    \centering
    \includegraphics[scale = 0.2]{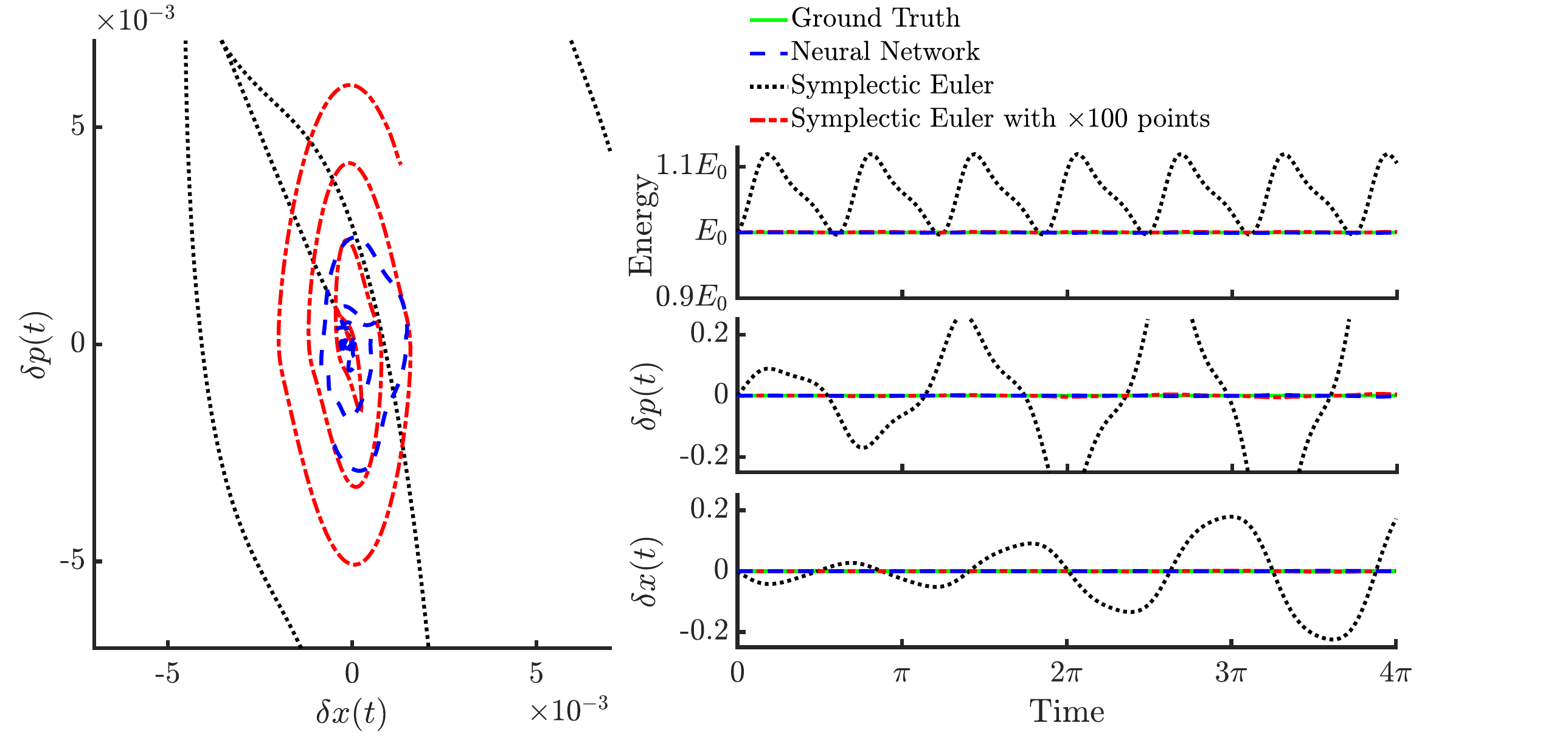}
    \caption{Comparing the ground truth (green) with the approximated solutions obtained by NN (blue) and by symplectic Euler integrator. The NN is trained over $K=200$ time points while the integrator is evaluated at $K$ (black) and $100\times K$ (red) points. Left: The phase space of the numerical error. Right:  The error evolution in position and momenta, and the total energy in time. }
    \label{fig:NLosc}
\end{figure}

The performance of the  Hamiltonian NN  after its training is represented in Fig. (\ref{fig:NLosc}) by the blue curve. In addition,  we use the DEs solver \texttt{odeint}  of the \texttt{scipy} python package \cite{scipy} to solve the  system (\ref{eq:eqMotOsc}) and consider  the obtained  numerical solutions as the ground truth.
{We note that  the solvers provided by \texttt{scipy}  have exemplary  error control and adaptivity leading to excellent solution trajectories.} 
For comparison purposes, we also utilize the symplectic Euler method described in  Eqs. (\ref{eq:EulerScheme_q}),(\ref{eq:EulerScheme_p}) to solve the  DEs (\ref{eq:eqMotOsc}), and compare the solutions with those obtained by our proposed symplectic NN.  
{We point out that the ground truth data and the solutions obtained by symplectic Euler method are exclusively used to assess the performance of  the NN predictions and never used for the NN optimization. Essentially, the  Hamiltonian NN   does  not  use  any  data generated  by  traditional  numerical  solvers.} In Fig.  \ref{fig:NLosc} we present results obtained by the solver (green lines), by the NN (blue line), and by the symplectic Euler integrator (black and red).
After the network optimization we get $\ell_\text{max} = 3.3\cdot10^{-3}$. The smallest singular value of the Hessian of Hamiltonian (\ref{eq:Hosc}) is $\sigma_\text{min}=1$. Subsequently,  Eq. (\ref{eq:dz}) yields for both $\delta x$ and $\delta p$ the upper bound error $ 3.3\cdot10^{-3}$. Interestingly enough, the symplectic Euler method needs $100\times K$ time points to approach this maximum error.
In the case of Euler's method,  we present in Fig. \ref{fig:NLosc} two numerical solutions: one with the same time points $K$ used in the NN training (black), and a second with 100 times more points (red). 
 The left graph in Fig. \ref{fig:NLosc} demonstrates the phase space for the numerical errors where we observe that the errors in the NN's solutions are in the same order with the error obtained by the symplectic Euler when 100 times more time points are used. On the right panel of Fig.  \ref{fig:NLosc} we present  $\delta x(t)$ and $\delta p(t)$ and the  the total energy as a function of time  calculated by using the numerical solutions in the Hamiltonian (\ref{eq:Hosc}). An important result of this exploration is that,  in contrast to the Euler integrator, the NN's  solutions conserve the total energy locally. This is a consequence of the fact that the solutions obtained by the symplectic NN conserve the correct Hamiltonian rather than a perturbed one, which is the case with the symplectic integrators. Therefore, in the context of the energy conservation task, the Hamiltonian NN outperforms the symplectic Euler integrator. 

\begin{figure}[h]
\includegraphics[scale=.2]{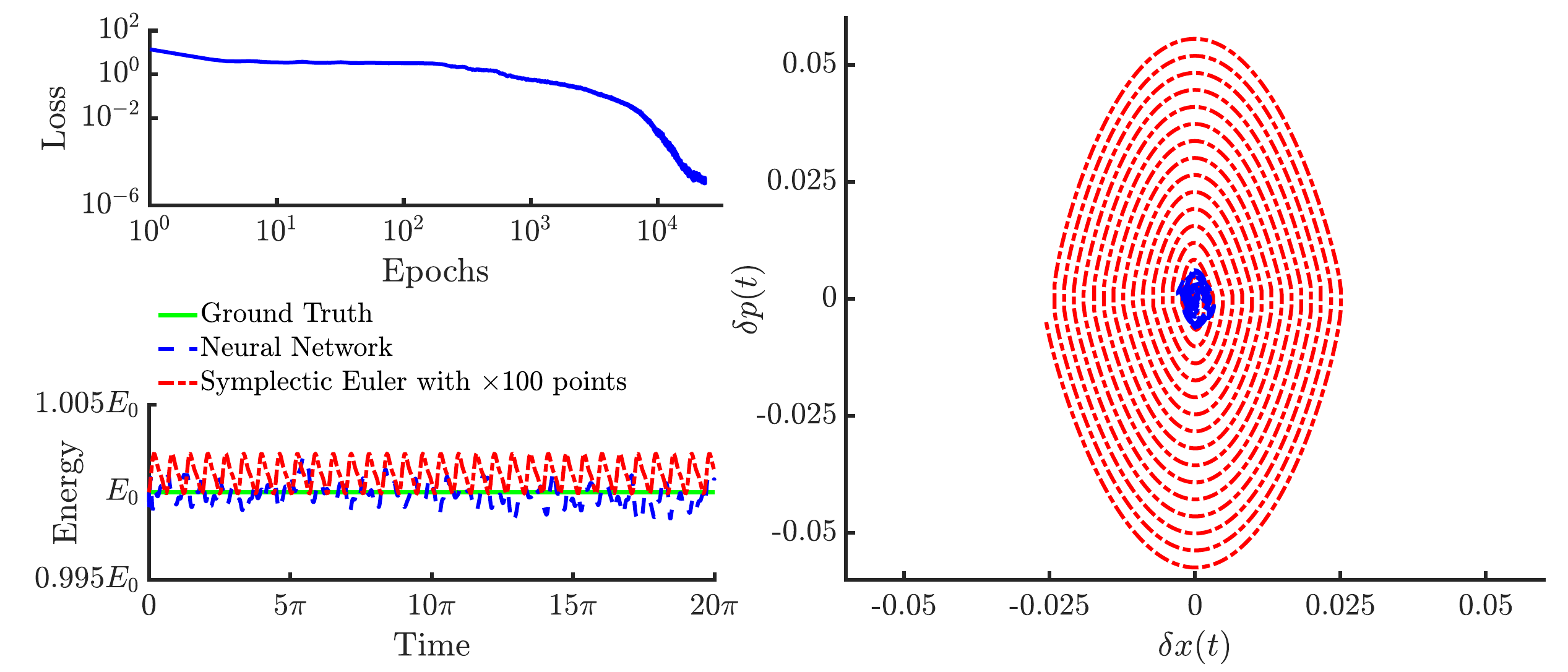}
\caption{Long term prediction. The upper left image demonstrates the loss function during the training of a Hamiltonian network. The lower left graph shows the ground truth energy  along with the predictions obtained by NN and by symplectic Euler method. The right panel presents the phase space of the numerical error for the predicted solutions.  }
\label{fig:longTermNLosc}
\end{figure}

{We validate the predictability of the Hamiltonian NN for long-term prediction by solving DEs in a longer time period. In particular, the system of Eqs. (\ref{eq:eqMotOsc}) is solved for the extended  time interval $[0, 20\pi]$ using the same initial conditions from the previous simulations, namely $(x_0,p_0)=(1.3,1.0)$. 
Although the previously used architecture provides solutions of high accuracy, we found that by using 80 neurons per hidden layer yields faster convergence in the NN optimization. Moreover, since the time interval is expanded the number of training points is increased accordingly to $K=500$ time points.
For the long-time prediction in this case, we use the regularization loss~\eqref{eq:Loss} with $\lambda=1$, which penalizes violations in the energy conservation.}
%For long time predictions it is efficient to use  the regularization  term $L_\text{reg}$ of Eq. (\ref{eq:Loss}) that penalizes any violation in the energy conservation law. }
%Specifically, for the nonlinear oscillator system we read Specifically, in the loss function of Eq. (\ref{eq:Loss_osc}) we add the regularization term } 
% \begin{align}
% \label{eq:regLossNL}
% {
% L_\text{reg} =  \frac{1}{K}\sum_{n=0}^K \left[ \left(\ham\left(\hat x^{(n)}, \hat p^{(n)}  \right)   -E_0 \right)^2 \right],
% }
% \end{align}
%{ with $\lambda=1$ and where $\ham$ is given by Eq. (\ref{eq:Hosc}). }
{The NN predictions are compared to solutions obtained by the symplectic Euler method using $5\times 10^4$ points. This represents 100 times more points than those used for the network optimization.  The results of the long time solutions are presented in Fig. \ref{fig:longTermNLosc}. The loss function during the training is shown by the left upper image in Fig. \ref{fig:longTermNLosc}. The lower panel represents the ground truth energy (green solid line), the energy obtained by the Hamiltonian NN (dashed blue), and the energy that the symplectic Euler method (red dashed-dotted) computes. We observe that the neural network conserves the energy slightly better than the numerical integrator. The right panel of Fig. \ref{fig:longTermNLosc} is the phase-space error, similar to Fig. \ref{fig:NLosc}, where we observe that the error obtained by the symplectic Euler (red dashed-dotted) constantly increases in time much faster than the error that we obtain from the NN solutions. Interestingly enough, we observe that although the energy is conserved comparably well by the two approaches, the Hamiltonian NN out-performs the symplectic Euler method in terms of the accuracy of the predicted solutions.  That happens because a NN solver simultaneously satisfies all the equations of DE system  and conserves the original Hamiltonian, whereas, integrators conserve a perturbed Hamiltonian accumulating errors in time. }

\subsection{Chaotic system}
%\label{sec:HH}
We demonstrate further the efficiency of the proposed symplectic NN by solving the equations  for a chaotic two-dimensional dynamical system. In particular, we solve the canonical equations for the  H\'{e}non-Heiles (HH) system \cite{HH_1964} that  describes the non-linear motion of a star around a galactic center with the motion restricted to a plane. The HH system has four degrees of freedom in the phase space where ${\bf z}=(\qq,\ppp)^T=\left(x,y,p_{x}, p_{y}\right)^T$.  The  Hamiltonian and the total energy of this system is 
\begin{align}
  \label{eq:HHpotential}
  \ham = \frac{1}{2}\left(p_x^2 +p_y^2 \right) + \frac{1}{2}\left(x^2 +y^2 \right) + \left( x^2y - \frac{y^3}{3} \right).
\end{align}
 The Hamilton's equations results in the nonlinear DEs system:
\begin{align}
\label{eq:HHq}
      \dot{x} &= p_{x},      &        \dot{y} &= p_{y},    \\ 
\label{eq:HHp}
    \dot{p}_{x} &= -\left( x + 2  x y \right), &  \dot{p}_{y} &= -\left( y +  x^{2} - y^{2}  \right).
\end{align}
For the HH system we are seeking  approximate solutions ${\bf \hat z}\in {\rm I\!R}^4$. Accordingly, we employ a fully connected feed-forward NN with four   outputs ${\bf N} \in {\rm I\!R}^4$ used to parametrize ${\bf \hat z}$ according to the general formula (\ref{eq:par_sol}).  The initial conditions for the numerical experiment are $\left(x_0,y_0,p_{x,0},p_{y,0}\right) = (0.3, -0.3, 0.3, 0.15)$,   corresponding to the  energy $E_{0} = 0.13$. The  maximal Lyapunov exponent   for this set of initial conditions is $\nu = 0.069$, and since  $\nu$ is positive,   the motion is chaotic  \cite{lyapExp}.  The network consists of two hidden layers with 50 neurons per hidden layer. An equally spaced grid of $K= 100$ is initialized in the time interval $t=[0,6\pi]$ that corresponds to 1.3 Lyapunov times. These points are used as the training set and  are perturbed in the beginning of every epoch by using a random term  obtained by a normal distribution with zero mean and with a standard deviation $0.18\pi$.   The loss function is defined by Eqs. (\ref{eq:HHq}), (\ref{eq:HHp}), and according to Eq. (\ref{eq:Loss}), as
\begin{align}
    \label{eq:Loss_HH}
    L &=  \frac{1}{K}\sum_{n=0}^K \left[ \left( \dot{\hat x}^{(n)} - \hat p_{x}^{(n)}  \right)^2 +    \left( \dot{\hat y}^{(n)} - \hat p_{y}^{(n)}  \right)^2 \right.   \nonumber \\
     &+  \left(\dot{\hat{p}}_{x}^{(n)} + \hat x^{(n)}  + 2\hat{x}^{(n)}\hat{y}^{(n)}\right)^{2} 
     \nonumber \\
   &+  \left.  \left(\dot{\hat{p}}_{y}^{(n)} + \hat y^{(n)} + \left(\hat{x}^{(n)}\right)^2 - \left(\hat{y}^{(n)}\right)^2\right)^2\right] .
\end{align}

\begin{figure}
    \centering
    \includegraphics[scale=0.25]{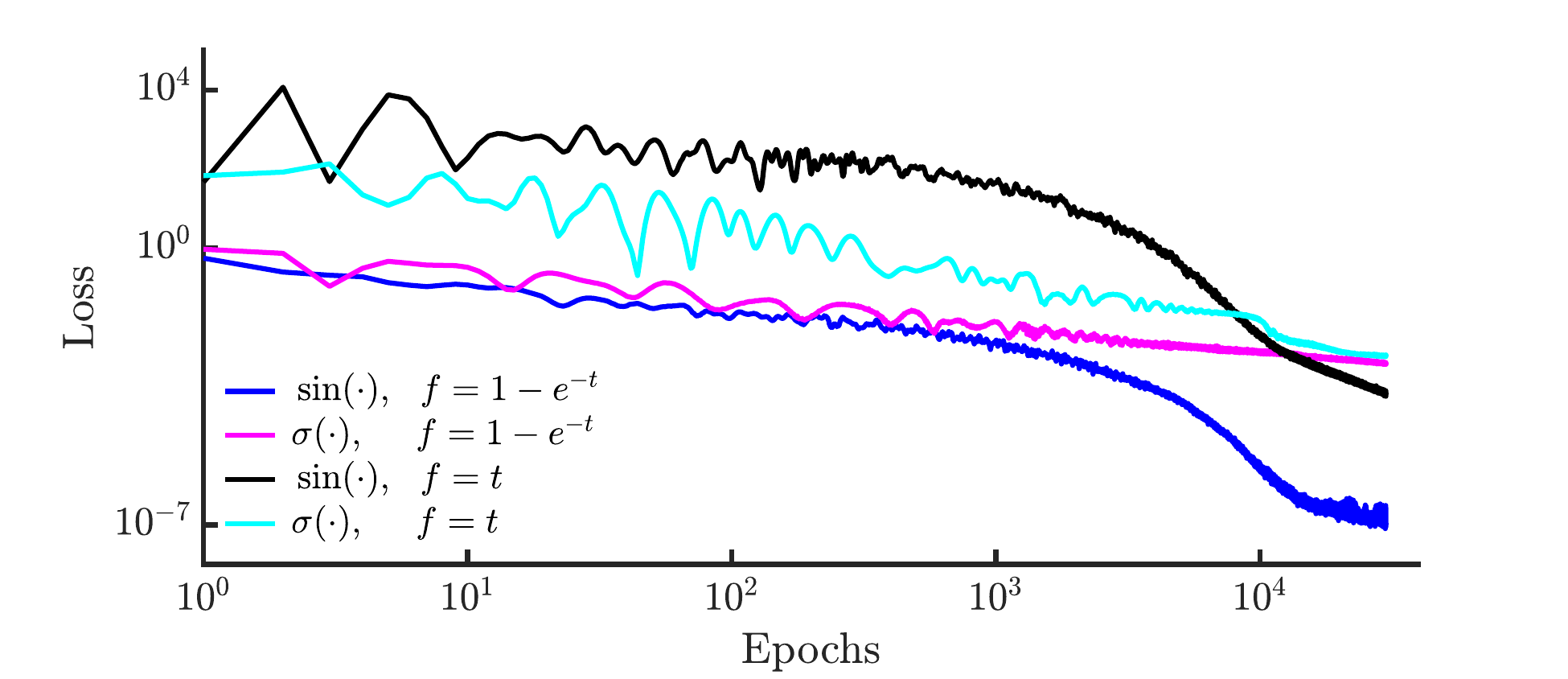}
    \caption{NN solves the equations of motion for the HH system. Loss function in log-scale during the training for a different combinations of activation and parametric functions $f$ shown by the legend.}
    \label{fig:HH_loss}
\end{figure}
\begin{figure}
    \centering
    \includegraphics[scale=0.2]{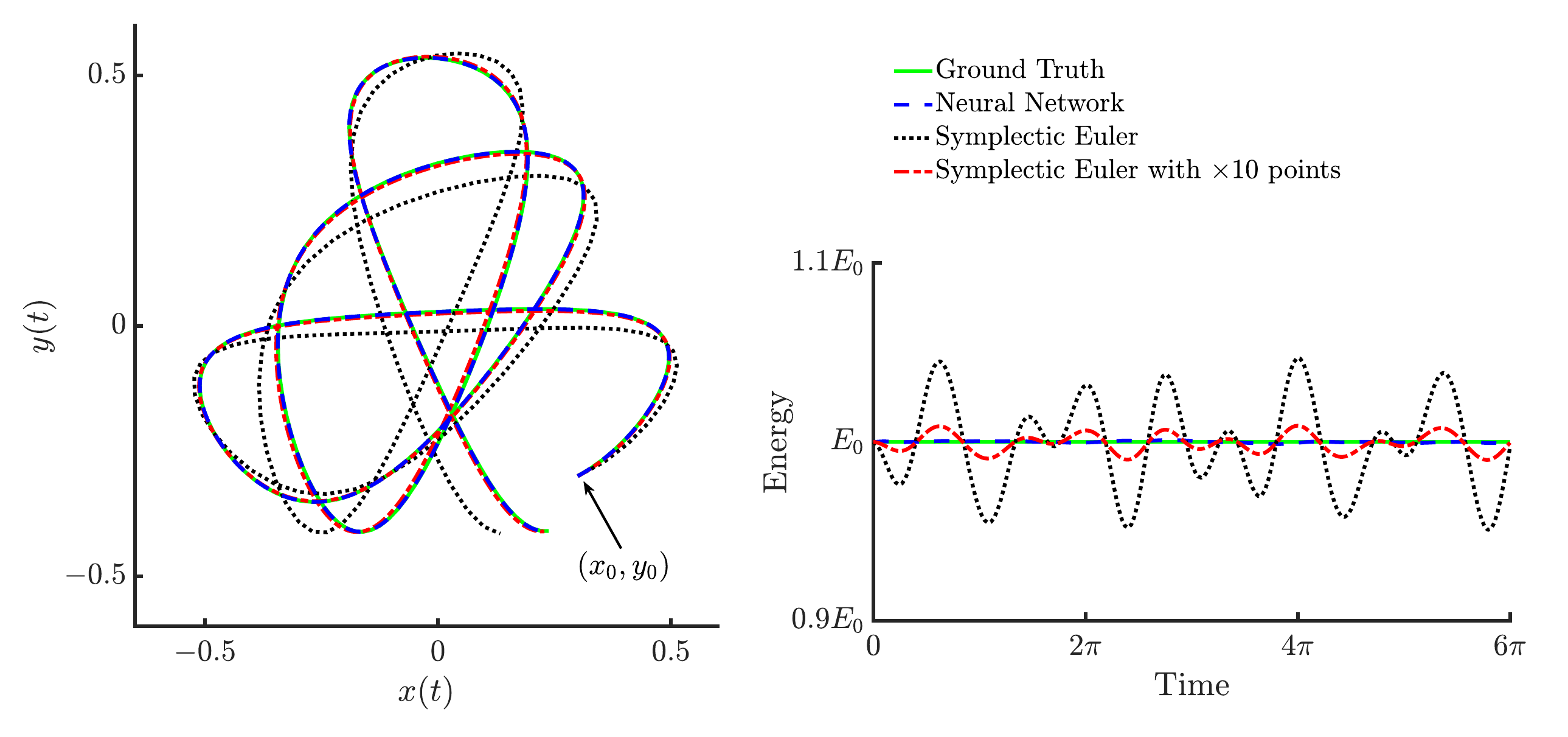}
    \caption{Left: The orbit for the HH system in the $x-y$ plane obtained by a NN (blue) that is trained in $K=100$ time points and by symplectic Euler integrator evaluated in $K$ (green) and $10\times K$ (orange) points. Red curves are considered as the ground truth and obtained by a numerical solver.     Right:  Energy of the HH system with time.  The Hamiltonian NN conserves energy locally while the symplectic Euler method does not maintain constant energy levels even at the highest resolution.}
    \label{fig:HHresults}
\end{figure}

We examine four different network architectures similar to  the nonlinear oscillator system, namely for different activation and parametric functions. The networks are trained for $3\cdot10^4$ epochs  by using Adam optimizer with learning rate $8\cdot 10^{-3}$.
After training for long enough to ensure convergence in the loss function we find this number of epochs is sufficient to optimize the network. 
%Moreover, we choose the specific learning rate after performing many experiments for different rates.  
In Fig. \ref{fig:HH_loss}, we show the loss function (\ref{eq:Loss_HH}) in the training where each color corresponds to a different architectures according to the legend in the figure. Again, the choice of $\sin(\cdot)$ activation and $f(t)=1-e^{-t}$  yields the best network performance. 
In Fig. \ref{fig:HHresults}, we compare the approximated trajectories and the energy obtained by  the symplectic NN (blue lines), and by a symplectic Euler integrator which is evaluated in $K$ and in $10\times K$ time points (shown by black and red lines, respectively). Solutions obtained by a solver are considered as the ground truth (green curves). The left panel in Fig.  \ref{fig:HHresults} shows the orbit in the $x-y$ plane  where the Hamiltonian NN solution is indistinguishable from the ground truth.
The right panel represents the total energy in time where the NN solutions conserve the energy better than the solutions obtained by the symplectic Euler method.
The symplectic Euler must use an order of magnitude higher resolution than NN to capture the correct orbit portrait, however, the energy is still not conserved locally.  
%We find, but do not show, in Fig. \ref{fig:HHresults} that the symplectic Euler requires two order of magnitude higher resolution in order to conserve the energy as well as the solutions obtained by the Hamiltonian network.

\begin{figure}[h]
    \centering
    \includegraphics[scale=0.21]{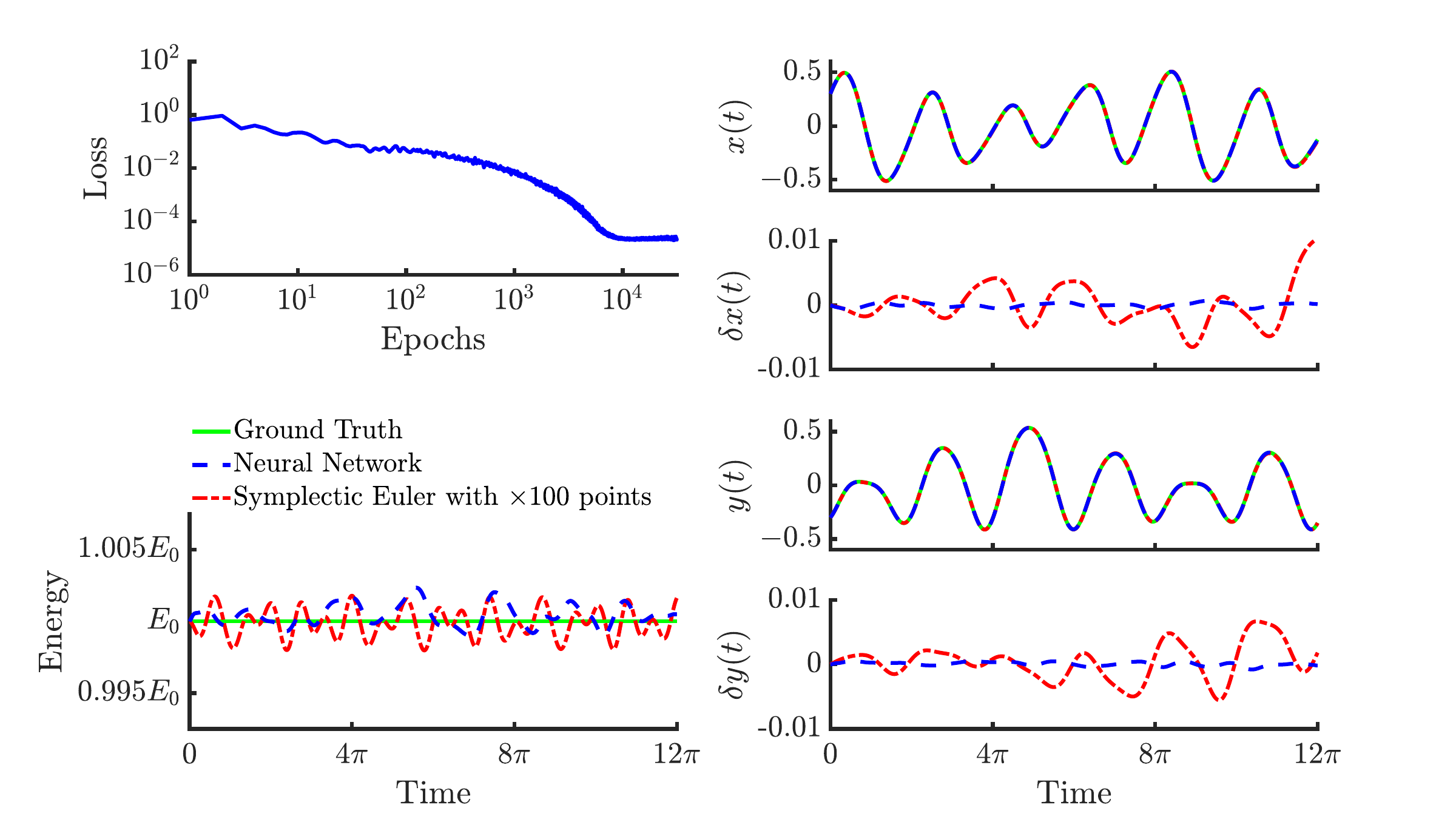}
    \caption{Prediction on 2.6 Lyapunov times for the HH system. The left upper graph indicates the training loss of a Hamiltonian network. The left lower plot is the total energy obrained by Hamiltonian NN and by symplectic Euler evaluated in $10\times$ more  than the training points. The right panels represent the predicted position states $x(t)$, $y(t)$ along with the associate numerical errors $\delta x(t)$ and $\delta y(t)$.    }
    \label{fig:HHnew}
\end{figure}

{ We extend the integration time for the HH system to $[0, 12\pi]$, which corresponds to \resub{$2.6$ Lyapunov times}. For the long-time prediction we employ the regularization term $L_\text{reg}$ of Eq. (\ref{eq:Loss}) with $\lambda=0.5$ }
%and with}
%\begin{align}
%\label{eq:regLossHH}
%{
%L_\text{reg} =  \frac{1}{K}\sum_{n=1}^K \left[ \left(\ham\left(\hat x^{(n)}, \hat y^{(n)}, \hat p_x^{(n)}, \hat p_y^{(n)}  \right)   -E_0 \right)^2 \right],
%}
%\end{align}
%{where $\ham$ is given by Eq. (\ref{eq:HHpotential}). }
{The network architecture consists of two hidden layers with 80 neurons per layer. The network optimization uses 500 time points. \resub{For the training of this model, we found that using  sequential learning \cite{nips2021pinns} is more efficient. First, we train the model for a short integration time range of $[0,6\pi]$ and save the network parameters; the network is trained for  $2\cdot 10^4$ epochs  with a learning rate of $8\cdot 10^{-3}$. Then, we load the previously saved parameters and train the model in a larger domain of $[0,12\pi$] for $5\cdot 10^4$ epochs and with  learning rate of $5\cdot 10^{-3}$. This  transfer learning application enhances the learning and, therefore, the network converges faster to the solutions than starting the training from random initialized parameters.} %
The results are demonstrated by Fig. \ref{fig:HHnew} where the left upper graph indicates the loss function during the training of the Hamiltonian network. For comparison,  we present the NN results in blue along with the solutions obtained by a symplectic Euler evaluated with $10\times$ more points than the training points. 
The lower plot in left  panel shows the energy where we observe that both the NN (blue) and the symplectic Euler (red) conserve the correct (green) energy with a error of about  the same order. The right panel of Fig. \ref{fig:HHnew} represents the predictions of the position state $x(t)$ and $y(t)$ along with the associated numerical error denoted by $\delta x$ and $\delta y$, respectively. As we observed in the nonlinear oscillator system,  the solutions obtained by the NN presents lower numerical error than the symplectic integrator, although both methods conserve the energy comparably well.}

%######################### 

\section{Conclusion}
In recent years, machine learning has made in-roads in traditional science and engineering fields. NNs have attracted scientists' interest  due to their outstanding capabilities in regression,  classification, and prediction tasks. Since these methods are relatively new to physics, there are many physical concepts that have not been  embedded yet in the  structure of NNs. In this work, we proposed a physics-inspired unsupervised NN for solving DEs that describe the temporal motion of  dynamical systems. 
The   Hamiltonian formulation  is embedded in the NN through the loss function and therefore, the predicted solutions conserve energy.
{The loss function is solely constructed by the network predictions and does not use any ground truth data. The   proposed method  does  not  use  any  data  generated  by  traditional  numerical  solvers. Hence, the proposed Hamiltonian network provides  a data-free unsupervised learning method. Although  the Hamiltonian network presented in the current work is an unsupervised model, the generalizations to the proposed network could incorporate data in a  semi-supervised fashion. Nevertheless, in this study we focused on the exploration of the baseline unsupervised model and leave the semi-supervised case for future work}.

  A smooth and bounded parametric  form of solutions was introduced in this study that makes the proposed architecture  a symplectic network, and subsequently, a time-invariant unit. By appropriately choosing the activation function a better domain knowledge is provided that drastically improves the network performance. Moreover, the proposed Hamiltonian architecture allows the network outputs to share their weights. Sharing the learning parameters helps the NN to discover underlying co-dependencies and subsequently, improves the network predictability in learning solutions that satisfy nonlinear systems of DEs.
  \resub{The Hamiltonian structure of the proposed NN allows the use of  a regularization term that penalizes violation in the energy conservation law. This penalty drastically improves the network performance especially for long time solutions. 
  The experiments presented in this work indicate  that in order to get accurate solutions for larger integration times,  more hidden neurons and time points are required  increasing  the network complexity and the computational cost. 
  This cost can be potentially  reduced  by parallelizing the calculations since each time point is treated independently, however, such an implementation is not presented in this study. In the limit of very long integration times, we expect to need very large network complexity and batches of time points,  thus, a parallel implementation will be crucial. }
  An error analysis was developed in this work which can be used to analyze how the errors in the predicted solutions propagate in time. In addition, this error analysis provides a threshold in the loss function, where we can early-stop  training the network when a certain accuracy occurs, namely a lower error in the predicted solutions is ensured.

There are several advantages in using NN solvers instead of traditional {symplectic} numerical integrators for solving DEs. The solutions obtained by a NN are continuous, smooth, and in an analytical form. Due to many outputs with shareable weights, the Hamiltonian NN discovers solutions that satisfy  the Hamilton equations simultaneously and consistently.  Subsequently, the NN solver conserves the correct Hamiltonian in contrary to symplectic integrators that conserve a slightly perturbed Hamiltonian. We  outlined that  the solutions obtained by the  NN conserve the energy  locally along with all the time points, and out-performs the symplectic Euler integrator that predicts an energy with a fluctuating error term. \resub{In addition to the first order  Euler method, there are higher order symplectic integrators that accumulate less error than semi-implicit Euler but with a larger computational cost.  Such a comparison between the proposed NN solver and higher order integrators is not presented in this study.}
In problems where energy conservation is crucial, the Hamiltonian NN will show better performance than symplectic integrators. 
Moreover, NN solvers can potentially    possess advantages over state of the art integrators such as the \texttt{odeint} from the \texttt{scipy} Python package.  As pointed out by \cite{lagaris1998}, the calculations for a NN can be efficiently implemented on parallel architectures leading to significant speed-up. This is possible because NN solvers evaluate the time points independently. In the years since that original work of    \cite{lagaris1998} appeared, hardware innovations such as GPUs have made the parallelization of NN  even more accessible. On the contrary, time-parallel algorithms for traditional numerical integrators are challenging to develop and implement since the computation at a time point requires solutions at prior time points. 
%An overview of recent advantages and challenges in parallel in time integration methods are summarized  by \cite{parallelDE2015}. Recently, it has been shown by \cite{parallelTraining2020} that modern  methods used to parallelize  the time integration in traditional solvers can be used for layer-parallel training of deep neural networks. \dls{I don't think you need to repeat this sentence here. You already have all this in the introduction.}
%
Additionally, as the number of the differential equations in a system increases the problem of the `curse of dimensionality' is observed  making the numerical integrators inefficient due to the rapidly increase of the computational cost. On the other hand, it has been shown by \cite{pnas2018, spiliopoulos2018} that the problem of the `curse of dimensionality' does not occur in neural network differential equations solvers. Subsequently, in high-dimensional problems such as many body problems, we expect the Hamiltonian NNs to out-perform  regular symplectic integrators.  
% Last but not least, recently it was shown by \cite{cedric2020} that a NN solver is able to learn solutions in a wide range (bundle) of initial conditions.  We are able to train a network to learn the general solution of differential system across many different initial conditions. This has the potential to significantly speed up applications in which many system solves would otherwise be required.}
%This is a great advantage of neural networks over any numerical integrator.}
%
Considering that  Hamiltonian formulation provides a solid framework for theoretical extension in many areas of physics such perturbation approaches and theory of chaos, as well as statistical and quantum mechanics, the  proposed Hamiltonian NN  provides fertile ground on which  modern research problems can potentially be handled.

% \resub{ {\bf MAYBE: }  Potentially, both methods can be combined in a deep architecture, where HNN can learn the Hamiltonian from data and then,  the proposed networks will forecast dynamical behavior by solving equations of motion replacing the embedded integrators. However, this proposal is out the goal of this work which focus only on the Hamiltonian neural network solver baseline.  }

% The Hamiltonian formulation provides a solid framework for theoretical extension in many areas of physics such perturbation approaches and theory of chaos, as well as statistical and quantum mechanics.  Hence, the  proposed Hamiltonian NN  provides fertile ground on which  modern research problems can potentially be handled.

\begin{acknowledgments}
This research did not receive any specific grant from funding agencies in the public, commercial, or not-for-profit sectors. The authors would like to thank   fruitful  discussions with Prof. E. Lagaris, Prof. G. P. Tsironis, and Prof. E. Kaxiras.
\end{acknowledgments}

% \bibliography{HNN_solver}

\end{document}